\documentclass[preprint,aps,showpacs,showkeys]{revtex4}
\newcommand{\sect}[1]{\setcounter{equation}{0}\section{#1}}
\newcommand{\bfm}[1]{\mbox{\boldmath${#1}$}}

\begin{document}
\draft
\title{PHASE SPACE CELL IN NONEXTENSIVE\\ CLASSICAL SYSTEMS}
\author{F. Quarati\footnote{Present address: Dept. of Physics and Astronomy, University of
Glasgow, Scotland, U.K.}}\email{francesco.quarati@ca.infn.it}
\affiliation{Dipartimento di Fisica - Universit\'a di Cagliari,
I-09042 Monserrato, Italy}
\author{P. Quarati}\email{piero.quarati@polito.it}
\affiliation{Dipartimento di Fisica - Politecnico di Torino,\\
Corso Duca degli Abruzzi 24, I-10129 Torino, Italy, and\\
Istituto Nazionale di Fisica Nucleare - Sezione di Cagliari,
I-09042 Monserrato,Italy}
\date{\today}

\begin {abstract}
We calculate the phase space volume $\Omega$ occupied by a
nonextensive system of $N$ classical particles described by an
equilibrium (or steady-state, or long-term stationary state of a
nonequilibrium system) distribution function, which slightly
deviates from Maxwell-Boltzmann (MB) distribution in the high
energy tail. We explicitly require that the number of accessible
microstates does not change respect to the extensive MB case. We
also derive, within a classical scheme, an analytical expression
of the elementary cell that can be seen as a macrocell, different
from the third power of Planck constant. Thermodynamic quantities
like entropy, chemical potential and free energy of a classical
ideal gas, depending on elementary cell, are evaluated.
Considering the fractional deviation from MB distribution we can
deduce a physical meaning of the nonextensive  parameter $q$ of
the Tsallis nonextensive thermostatistics in terms of particle
correlation functions (valid at least in the case, discussed in
this work, of small
deviations from MB standard case).\\
\end {abstract}
\pacs{05.20.-y, 05.70.-a} \keywords{Classical Statistical
Mechanics, Thermodynamics} \maketitle
%%%%%%%%%%%%%%%%%%%%%%%%%%%%%%%%%%%%%%%%%%%%%%%%%%%%%%%%%%%%%%%%%%%%%%

\sect{Introduction}

Statistical description of a system of $N$ particles requires the
subdivision of the phase space into equidimensional elementary
cells of phase volume $\Delta\,\Omega$, which can be determined by
the laws of nature (comparison with quantum evaluation of energy
state density) and experimentally measured, for instance, in the
low temperature heat capacity of a crystal or in the
Stefan-Boltzmann constant.\\ In the phase space volume of a system
of particles described by quantum distribution the smallest
elementary cell is the third power of the Planck constant. For
classical particles the elementary cell is, in principle,
undetermined. This is true particularly in the limiting case of
small occupation numbers (when MB distribution is valid) and the
phase space volume of a cell acquires arbitrary values.\\ The
problem of dividing the phase space into finite cells was solved
when a natural way to derive dimension of elementary cells within
MB statistics using energy quantization was found. However,
quantum discontinuity can be lost when dimension of the volume
containing the gas increases and the quantum states become even
and even more numerous \cite{1}. Only the insertion of the Pauli
exclusion principle enables us to solve this problem.\\
Boltzmann's request of a great number of particles into the cell
is usually not verified (one particle every 30000 cells in normal
conditions)\cite{2}. However, if the number is small, it is
possible to combine many cells and form a greater cell (macrocell)
containing more particles \cite{3}.\\

This paper aims at examing first of all how , in the nonextensive
thermostatistics (NETS), the elementary cell differs from the one
of the extensive MB case, requiring explicitly that the number of
accessible microstates be the same in both (extensive and
nonextensive) phase spaces and obtain also, within a classical
scheme and without quantum arguments, explicit expressions of the
cells. Let us briefly recall that NETS has been developed in the
last years mainly after the works of Tsallis \cite{4,5} which
introduced a generalised entropy, featured by an entropic
parameter $q$, whose meaning is not fully understood yet (more
generally, it can be defined in terms of fluctuations of
intensive quantities like temperature \cite{6,7,8}; explicit
analytical expressions of $q$ in terms of physical quantities
exist in few cases, like, for instance, in turbulence problems
\cite{9} and in the description of solar plasma \cite{10,11}).\\
The value of $q$ is not easily given a priori for a given
particular system, without comparing calculations to available
experimental results.\\ The NETS is now applied to a great
variety of problems, from genetics and biology  to astrophysics
(see ref.s \cite{12} and \cite{13} for a complete list of basic
works and applications).\\ The approach to NETS we are showing in
this paper, is based on the analysis of deviations from standard
phase space volume and on a new definition of elementary cell.
This new approach will probabily provide, in the near future, a
better understanding of the meaning of the parameter $q$. The
validity of our treatment can be extended to statistics different
from the Tsallis version of NETS, when complete deformed
distribution functions are considered, and to
large deviations from MB phase space volume.\\
In Section 2, by comparing the number of allowed microstates
$W_0$ in the MB extensive phase space $\Omega_0$ and in the
deformed phase space $\Omega$, we derive the dimensions of both
elementary cells $\Delta\Omega_0$ and $\Delta\Omega$. The cell
$\Delta\Omega$ results to be smaller than $\Delta\Omega_0$ (if
$q<1$) so that, in this case, the third power of Planck constant
is not suitable as the value of $\Delta\Omega_0$ because $h^3$ is
the smallest cell admissible due to the Heisenberg principle.
Otherwise $\Delta\Omega$ is larger than, $\Delta\Omega_0$ if $q>1$
($h$ is expressed in units [energy length]).\\ By considering the
nonextensive Tsallis equilibrium (or steady state, or long-term
stationary state of a nonequilibrium system) distribution, we
limit ourselves to small deviations, with the absolute value of
the parameter $\delta=(1-q)/2$ much smaller than one (we recall
that
for $q\rightarrow1$ all the MB results must be recovered).\\
We must realize that deviations, although small, are not
negligible in those energy islands of $\Omega$ where quantities
like, for instance, nuclear reaction rates, rates of atomic
processes and of chemical reactions, electron transport in
semiconductors are very sensible to them \cite{10,14,15,16}. In
Section 3, we discuss the case of nonextensive classical ideal
gas. We compare our results with those recently derived by several
authors, among them see ref.s \cite{17,18,19}, and we calculate
the entropy, the chemical potential and free energy discussing the
results. We deduce also in Section 4 an interpretation of the
parameter $q$ using the calculated phase volume $\Omega$ as a
function of fractional deviation, occupied by a nonextensive
system of $N$ classical particles. The physical meaning is
derived in terms of particle correlation function. Conclusions
are reported in Section 5.

%%%%%%%%%%%%%%%%%%%%%%%%%%%%%%%%%%%%%%%%%%%%%%%%%%%%%%%%%%%%%%%%%%%

\sect{Phase space volume and elementary cell}

The state of a system of $N$ particles is specified in the $6\,N$
dimensional phase space ($\Gamma$ space). Its volume $\Omega[n_r]$
can be divided into small cells of volume $\Delta\Omega$, so that
coordinates do not vary sensibly within them.\\ The volume
contains the $N$ particles distributed in a certain set of numbers
$(n_1,\,n_2,\,\cdots,\,n_t)=[n_r]$ and is given by:
\begin{eqnarray}
\Omega[n_r]=\frac{N!}{n_1!\,n_2!\,\cdots,\,n_t!}\,(\Delta\Omega)^N
\ .\label{1}
\end{eqnarray}
The number of accessible microstates is defined (including the
factor $1/N!$) by \cite{20}
\begin{eqnarray}
W=\frac{\Omega}{N!\,(\Delta\Omega)^N} \ .
\end{eqnarray}
It is well known that $\ln\Omega$ can be represented through the
Stirling approximation
\begin{eqnarray}
\ln\Omega\approx N\,\ln N+N\,\ln\Delta\Omega-\sum_in_i\ln n_i \ .
\end{eqnarray}
Having $\Omega$ a maximum, its logarithm can be expanded in a
series of powers of $\partial n_i=n_i-n_{i_0}$, where $n_{i_0}$
is the distribution function for which $\Omega$ has a maximum,
indicated by $\Omega_0$ (the symbol $\partial$ means variation).
After using the Lagrange method with the usual constraints and
with negligible interactions the following is obtained:
\begin{eqnarray}
\nonumber &&n_{i_0}=A_{\rm M}\,e^{-x_i} \
,\hspace{2cm}\sum_in_{i_0}=N \
,\hspace{2cm}\sum_in_{i_0}\,x_i={3\over2}\,N \ ,\\
&&x_i=\beta\,\epsilon_i \ ,\hspace{2cm}\beta={1\over k\,T} \ ,\\
\nonumber
&&A_{\rm
M}=\frac{N}{V}\,\left(\frac{\beta}{2\,\pi\,m}\right)^{3/2}\,\Delta\Omega_0
\ .
\end{eqnarray}
The mass of the particles $m$ is in energy units. The number of
accessible microstates in $\Omega_0$ is of course
\begin{eqnarray}
W_0=\frac{\Omega_0}{N!\,(\Delta\Omega_0)^N} \ ,
\end{eqnarray}
(let consider the following example: for a classical ideal gas, if
$N$
is large we have that $W_0=e^N/\sqrt{2\,\pi\,N}$, see Sect.3).\\
We are now interested in the volume in the $\Gamma$ space
corresponding to a NETS distribution having small deviations from
the MB distribution. The deviations in the high energy tail are
particularly interesting. Following the approach to general
statistical problem in physics by Bohm and Sch\"utzer \cite{21}
after expanding up to the second power in $\partial n_i$ the
Lagrange equation, alternative definition of $\Omega$ is obtained:
\begin{eqnarray}
\ln\Omega=\ln\Omega_0-{1\over2}\,\sum_i\frac{(\partial
n_i)^2}{n_{i_0}} \ ,
\end{eqnarray}
or
\begin{eqnarray}
\Omega=\Omega_0\,\exp\left[-{1\over2}\,\sum_in_{i_0}\,(\partial
\,f_i)^2\right] \ ,
\end{eqnarray}
with the fractional deviation from MB distribution given by
\begin{eqnarray}
\partial\,f_i=\frac{\partial\,n_i}{n_{i_0}} \ .
\end{eqnarray}
Let us introduce the non-Maxwellian distribution
\begin{eqnarray}
n_i=A_{\rm T}\,A_\delta\,\exp\left(-y_i-\delta\,y_i^2\right) \
.\label{7}
\end{eqnarray}
Eq.(\ref{7}) represents a distribution, mainly differing from
Maxwellian in the high energy tail, depending on the sign of
$\delta$ [$\delta>0,\,\,(q<1)$: depleted tail;
$\delta<0,\,\,(q>1)$: enhanced tail]; it can be derived, for
instance, from the Tsallis distribution
\begin{eqnarray}
n_{\rm T}=Z_q(\beta^\ast)^{-1}\,\left[1-(1-q)\,\beta^\ast\,
\left(\sum_i\frac{p_i^2}{2\,m}-U_q\right)\right]^{\frac{1}{1-q}}
\ ,\label{8}
\end{eqnarray}
when deviations from MB distribution are small, $Z_q(\beta^\ast)$
is the generalized partition function and $U_q$ the internal
energy. Eq. (\ref{7}) may represent an equilibrium distribution
or a steady state, or a long-term stationary state of a
nonequilibrium system. In Eq.(\ref{7})
\begin{eqnarray}
A_{\rm
T}=\frac{N}{V}\,\left(\frac{\beta^\ast}{2\,\pi\,m}\right)^{3/2}\,\Delta\Omega
\ .
\end{eqnarray}
$\Delta\Omega$ is the deformed elementary cell of the phase space
$\Omega$, $\beta^\ast$ is the Lagrange multiplier of the phase
space $\Omega$, different from $\beta$ (the Lagrange multiplier
in $\Omega_0$), defined below in (\ref{21}).\\
We also have
\begin{eqnarray}
&&A_\delta=1+{15\over4}\,\delta-30\,\delta^2 \ ,\\
&&y_i=\beta^\ast\,\epsilon_i \ .
\end{eqnarray}
$\epsilon_i$ are the energies of the different microstates with
the same values of those of the phase space $\Omega_0$. The
constraints are
\begin{eqnarray}
\sum_i n_i=N \ ,\hspace{2cm}\sum_in_i\,y_i={3\over2}\,N \ ,
\end{eqnarray}
where the second one is the energy NETS average value when
deviations are small. Other distributions, based on other
statistics can be used \cite{21bis,21tris}; of course, when
deformations are small, these distributions can be fitted with
good approximation by Eq. (\ref{7}). The use of exact
distributions, derived from other generalized statistics is
possible at least numerically; in this work, however, we limit
ourselves to consider only the distribution of Eq. (\ref{7}), in
order to simplify the question.\\ The volumes $\Omega_0$ and
$\Omega$ are different in size. $\Delta\Omega_0$ and
$\Delta\Omega$ are also different. We realize that the
differences could be considered negligible, while their effects
seem to be quite important for the evaluation of several physical
quantities. Let us remark that the number of microstates or
discrete events does not change from one space to the other one.
Therefore, we set the equation $W=W_0$, because we want to count
the same number of microstates both in $\Omega$ and $\Omega_0$.
Using the relations reported above in this Section , we can
explicitate the expressions of $\Omega$ and $\Delta\Omega$ after
simple calculations. After using the following relations:
\begin{eqnarray}
&&\frac{A_{\rm T}}{A_{\rm
M}}=\frac{\Delta\Omega}{\Delta\Omega_0}\,\left(\frac{\beta^\ast}{\beta}
\right)^{3/2}=\frac{e^{9\,\delta/4}}{A_\delta} \ ,\\
&&\frac{A_\delta^2}{A_{2\,\delta}}=1+\frac{1185}{16}\,\delta^2 \ ,\\
&&\sum_i\frac{n_i^2}{n_{i_0}}=N\,\left(1+18\,\delta-\frac{3799}{32}\,\delta^2\right)
\ ,\label{15}
\end{eqnarray}
noting that the explicit calculation of (\ref{15}) requires the
evaluation of the average value
\begin{eqnarray}
\left\langle\frac{e^{-x_i}}{e^{-y_i}}\right\rangle\simeq\exp\left(\frac{37}{2}
\,\delta-\frac{11841}{64}\,\delta^2\right) \ ,
\end{eqnarray}
finally, we obtain $\Omega$ and $\Delta\Omega$ as functions of
$\Omega_0$ and $\Delta\Omega_0$, respectively as follows:
\begin{eqnarray}
&&\frac{\Omega}{\Omega_0}=\exp\left[-{1\over2}\,N\,\left({1\over
N}\,\sum_i\frac{n_i^2}{n_{i_0}}-1\right)\right]=\exp
\left[-{1\over2}\,N\,\left(18\,\delta-\frac{3799}{32}\,\delta^2\right)\right]
\ ,\\
&&\frac{\Delta\Omega}{\Delta\Omega_0}=\exp\left[-{1\over2}\,\left(18\,\delta-\frac{3799}{32}\,\delta^2\right)\right]
\ .\label{17}
\end{eqnarray}
The systems featured by $\Delta>0$ ($q<1$) have $\Delta\Omega$
smaller than $\Delta\Omega_0$.\\
Therefore, we are not allowed in their classical extensive
description to take $\Delta\Omega_0=h^3$ because $h^3$ should be
the smallest elementary permissible cell.\\ On the other hand, we
shall verify that $\Delta\Omega_0$ must always be much larger than
$h^3$. It must be a macrocell.\\ The equation of state calculated
by means of the distribution (\ref{7}) is given by
\begin{eqnarray}
P\,V=N\,k\,T\,C_\delta=N\,C_\delta/\beta \ ,\label{18}
\end{eqnarray}
where
\begin{eqnarray}
C_\delta=1-5\,\delta+46\,\delta^2 \ .\label{19}
\end{eqnarray}
Let us anticipate that for a nonextensive classical ideal gas
after exact calculations with the exact distribution (\ref{8}),
one can obtain the following relation, at any value of $q$
\cite{17,18}
\begin{eqnarray}
&&P\,V=N\,\beta^\ast=N\,C_q/\beta \ ,\label{20}\\
&&\beta^\ast\,C_q=\beta \ ,\label{21}
\end{eqnarray}
where $C_q=Z_q(\beta)^{1-q}$.\\
We compare (\ref{18}) and (\ref{19}) to (\ref{20}) and (\ref{21})
in the case of small deviations. With the condition
$3\,N\,(1-q)/2=3\,N\,\delta>1$ the quantity $C_q$ reduces to
$C_\delta$ if we assume an elementary cell given by
\begin{eqnarray}
\Delta\Omega_0=\left(\frac{2\,\pi\,m}{\beta}\right)^{3/2}\frac{V}{N}
\ ,\label{22}
\end{eqnarray}
and we recover the correct formal expression of the equation of
state reported in Eq. (\ref{18}). This value of $\Delta\Omega_0$
imposes that
\begin{eqnarray}
A_{\rm M}=1 \ ,\hspace{2cm}{\rm and}\hspace{2cm}A_{\rm
T}=1-{3\over2}\,\delta+\frac{1221}{32}\,\delta^2 \ .
\end{eqnarray}
Therefore,  the requirements that $W=W_0$ and that the equation
of  state for NETS  classical systems be correctly expressed also
in the small deviations limit imply  that the standard phase
space elementary cell be given by the expression
(\ref{22}) (we send to Sect. 3 for some more details).\\
This elementary cell can indeed be considered a macrocell,
particularly if compared to the value of $h^3$, usually taken as
elementary cell. However, this requirement is not a problem,
because of the uncertainty of the classical elementary cell and
because Darwin and Fowler \cite{3} showed that macrocells should
be used to satisfy Boltzmann requirements of a great average
number of particles in each cell.\\ To take $\Delta\Omega_0$ of
Eq. (\ref{22}) means to have one particle in each macrocell and,
posing $\Delta\Omega_0=X\,h^3$, to have $1/X$ particles in each
microcell $h^3$, i.e. one particle in thousands of cells, where
\begin{eqnarray}
X=\left(\frac{2\,\pi\,m}{\beta}\right)^{3/2}\,\frac{V}{N}\,\frac{1}{h^3}
\ ,
\end{eqnarray}
which is not a pure number, but depends on $\beta$.\\
The elementary cell in the deformed phase space is from (\ref{17})
\begin{eqnarray}
\Delta\Omega=\left(\frac{2\,\pi\,m}{\beta^\ast}\right)^{3/2}\,\frac{V}{N}=
\left(\frac{2\,\pi\,m}{\beta}\right)^{3/2}\,\frac{V}{N}\,C_\delta^{3/2}
\ ,\label{24}
\end{eqnarray}
which is a macrocell. We can write $\Delta\Omega=Y\,h^3$, each
macrocell containing one of the $N$ particles, where
\begin{eqnarray}
\nonumber
Y=\left(\frac{2\,\pi\,m}{\beta}\right)^{3/2}\,\frac{V}{N}\,\frac{C_\delta^{3/2}}{h^3}
\ ,
\end{eqnarray}
is a quantity depending on $\beta$ and $q$ or on $\beta^\ast$.\\
The subdivision of the phase space volume $\Omega$ in microcells
equal to $h^3$ does not allow the conservation of the value of
the number of permissible microstates, nor the subdivision in
a fixed number of microcells independent on $\beta$ and $q$.\\
Instead of having an elementary cell which does not depend on the
Lagrange multiplier $\beta$ and is the third power of a universal
constant (Planck constant) and the number of elementary cells
forming the total phase space volume depending on the Lagrange
multiplier $\beta$, we have an elementary cell which depends on
the Lagrange multiplier $\beta$ (it is a macrocell). We also have
the number of cells depending only on the number of particles $N$
(with these positions we may satisfy the requirements to conserve
the number of accessible microstates both in extensive and in
nonextensive phase volumes $\Omega_0$ and $\Omega$ and to
preserve the correct form of the equation of state).
%%%%%%%%%%%%%%%%%%%%%%%%%%%%%%%%%%%%%%%%%%%%%%%%%%%%%%%%%%%%%%%%%%%%%

\sect{Application to classical  ideal gas: entropy, chemical
potential, free energy.}

Classical ideal gas model based on nonextensive thermostatistics
relations has been the subject of several studies since the first
applications of NETS \cite{22,23}. Classical ideal gas is
described by an unperturbed state of a system with long-range
interaction and the model can be solved analytically \cite{17}. In
the past the equation of state has been derived within several
schemes along the steps of evolution of NETS: non-normalized
\cite{22,23}, normalized \cite{17},  OLM (optimal Lagrange
multiplier) \cite{18}, incomplete statistics \cite{19,24,24bis},
among others.\\ In this Section we show synthetically that the
formally correct equation of state (\ref{20}) obtained  by means
of the exact expression of the distribution (\ref{8}), can be
deduced also in the small deviations case, by taking into account
the distribution (\ref{7}) if the macrocells of Eq. (\ref{22})
and of Eq. (\ref{24}) are taken as elementary cells.\\ The
equation of state specific for classical ideal gas remains form
invariant under nonextensive generalization of thermodynamics.
That is, it is valid for all $q$: $P\,V=N/\beta^\ast$. This means
that the equation of state of a NETS classical ideal gas is
formally equal to that of extensive statistical mechanics:
$P\,V=N/\beta$.\\ The Lagrange multiplier $\beta^\ast$ previously
introduced, associated with the constraint in NETS, is defined by
\begin{eqnarray}
\beta^\ast={1\over k_{\rm T}\,T_{\rm phys}} \ ,
\end{eqnarray}
$k_{\rm T}$ is a constant depending on $q$ which becomes the
Boltzmann constant $k$ for $q\rightarrow1$ \cite{18,25} and
$T_{\rm phys}$, the physical temperature, is
\begin{eqnarray}
T_{\rm phys}=\frac{C_q}{k_{\rm
T}\,\beta}=\frac{Z_q(\beta)^{1-q}}{k_{\rm
T}\,\beta}=\left(1+\frac{1-q}{k_{\rm T}}\,S_q^{\rm
T}\right)\,{1\over k_{\rm T}\,\beta} \ ,
\end{eqnarray}
($S_q^{\rm T}$ is the Tsallis entropy defined below in
Eq.(\ref{34}); in the first treatment of classical ideal gas
based on OLM approach \cite{25} the dependence on $q$ was
attributed only to $k_{\rm T}$ and not to $T_{\rm phys}$ or
$\beta^\ast$). The partition function $Z_q(\beta)$ is defined by
\begin{eqnarray}
Z_q(\beta)=\frac{V^N}{N!\,(\Delta\Omega_0)^N}\,\int\prod_i\,d^3p_i\,n_{\rm
T} \ ,
\end{eqnarray}
where $n_{\rm T}$ is the square bracket factor of Tsallis distribution [Eq.(\ref{8})].\\
Two other quantities that are useful for NETS classical ideal gas
calculations are: the coefficient
\begin{eqnarray}
C_q=\frac{V^N}{N!\,(\Delta\Omega_0)^N}\,\int\prod_i\,d^3p_i\,\frac{n_{\rm
T}^q}{Z_q(\beta)^q}=Z_q(\beta)^{1-q} \ ,
\end{eqnarray}
and the internal energy
\begin{eqnarray}
U_q={1\over
C_q}\,\frac{V^N}{N!\,(\Delta\Omega_0)^N}\,\int\prod_i\,d^3p_i\,\sum_j\frac{p_j^2}{2\,m}\,\frac{n_{\rm
T}^q}{Z_q(\beta)^q} \ .
\end{eqnarray}
From the explicit expression of $Z_q$, $C_q$ and $U_q$ (we do not
report it here and we send the reader to ref.s \cite{17,18}), we
may verify that the  above three functions $Z_q$, $C_q$ and $U_q$
depend (when $q\not=1$) on the elementary cell $\Delta\Omega_0$ as
\begin{eqnarray}
\nonumber && Z_q(\beta)\approx(\Delta\Omega_o)^{-N/(1-Q)} \ ,\\
&&C_q\approx (\Delta\Omega_o)^{-N\,(1-q)/(1-Q)} \ ,\\ \nonumber
&&U_q\approx (\Delta\Omega_o)^{-N\,(1-q)/(1-Q)} \ ,
\end{eqnarray}
where $Q=3\,N\,(1-q)/2$.\\
As well known, when $q\rightarrow1$ we have that
$Z_1\simeq(\Delta\Omega_0)^{-N}$ and $C_1$ and $U_1$ do not
depend on $\Delta\Omega_0$. The above functions enter into the
calculation of the equation of state, which can be derived by
means of the usual thermodynamic relations. As we have already
discussed in the previous Section, by considering a nonextensive
classical ideal gas with distribution function $n_i$ of  Eq.
(\ref{7}) (small deviations from MB distribution), we have
calculated that the equation of state is given by
$P\,V=N\,k\,T\,(1-5\,\delta+46\,\delta^2)$ i.e., in the limit of
small deviations we must have $\beta^\ast=\beta/C_\delta$ as it can be easily verified.\\
In fact, within the treatment illustrated in the previous
Section, we obtain that in the limit of small deviations and for
$3\,N\,\delta>1$ the expression of $C_q$ reduces to $C_\delta$ if
the elementary
cell (macrocell) $\Delta\Omega_0$ of Eq. (\ref{22}) is assumed.\\
With the expression of $\Delta\Omega$ given by (\ref{24}) in place
of $\Delta\Omega_0$ given by (\ref{22}) we obtain that $Z_q$,
$C_q$ and $U_q$ do not explicitly depend on the elementary cell.
$Z_q$ and $C_q$ do not depend on $\beta$ either:
\begin{eqnarray}
Z_q=\frac{\Gamma\left(\frac{2-q}{1-q}\right)}{\Gamma\left(\frac{2-q}{1-q}+{3\over2}\,N\right)}\,
\frac{N^N}{N!}\,\left(\frac{1}{1-q}\right)^{3\,N/2}\,\left[1+(1-q)\,{3\over2}\,N\right]^{\frac{1}{1-q}+
{3\over2}\,N} \ ,
\end{eqnarray}
and $C_q$ and $U_q$ can be calculated from the relations
\begin{eqnarray}
C_q=Z_q^{1-q} \ ,\hspace{2cm} U_q={3\over2}\,N\,\frac{C_q}{\beta}
\ ,
\end{eqnarray}
(the functions $\Gamma(x)$ can be calculated by means of the
relation $\Gamma(x)=\sqrt{2\,\pi}\,x^{x-1/2}\,e^{-x}$).\\
Let us note from Eq. (\ref{1}) that space phase volumes $\Omega_0$
and $\Omega$ depend on the elementary cells $\Delta\Omega_0$ and
$\Delta\Omega$, respectively. If $\Delta\Omega_0$ is a constant,
like $h^3$, $\Omega_0$ does not depend on $\beta$, but only on $N$
and $\Omega$ depends on $N$ and $q$. Instead, if $\Delta\Omega_0$
has the expression of Eq. (\ref{22}) and $\Delta\Omega$ is given
by Eq. (\ref{24}), then $\Omega_0$ is also a function of
$\beta$ and $\Omega$ is also a function of $\beta$ and $q$.\\
Let us now derive one of the thermodynamics quantities depending
on $\Delta\Omega$, the entropy.\\ The Boltzmann entropy is defined
by
\begin{eqnarray}
\nonumber S_{\rm B}=k\,\ln
W_0&=&{5\over2}\,N\,k+const.\\
&=&{5\over2}\,N\,k+N\,k\,\ln\left[\frac{V}{N}\,
\left(\frac{2\,\pi\,m}{\beta}\right)^{3/2}\,{1\over(\Delta\Omega_0)}\right]
\ ,
\end{eqnarray}
where $V\,(2\,\pi\,m/\beta)^{3/2}/\Delta\Omega_0$, is the single
particle number of cells.\\ Taking the elementary cell
$\Delta\Omega_0$ given by Eq. (\ref{22}), we obtain $S_{\rm
B}=5\,N\,k/2$ because the single particle number of cells equals
$N$ and the value of the constant is zero.\\ The nonextensive
entropy is \cite{4}
\begin{eqnarray}
\nonumber S_q^{\rm
T}=\frac{k}{1-q}\,\left(W^{1-q}-1\right)&=&\frac{k}{1-q}\,\left(W_0^{1-q}-1\right)\\
&=&\frac{k}{1-q}\left\{\exp\left[\frac{S_{\rm
B}}{k}\,(1-q)\right]-1\right\} \ ,\label{34}
\end{eqnarray}
where the condition $W=W_0$ has been used and therefore
\begin{eqnarray}
S_q^{\rm T}={5\over2}\,N\,k+(1-q)\,\frac{25}{8}\,N^2\,k=S_{\rm
B}\,\left(1+{5\over2}\,\delta N\right) \ .\label{311}
\end{eqnarray}
This relation shows that $S_q^{\rm T}$ tends to change as $N^2$
with large $N$. The chemical potential $\mu$ of a classical ideal
gas is an intensive quantity, defined by
\begin{eqnarray}
\mu=k\,T\,\ln\left[\frac{N}{V}\,\frac{\Delta\,\Omega_0}{(2\,\pi\,m\,k\,T)^{3/2}}\right]
\ ,\label{36}
\end{eqnarray}
where $\Delta\Omega_0$ is the elementary cell taken usually equal
to $h^3$. The quantity $\mu$ can be negative or positive and is
equal  to zero only at the particular value of temperature
\begin{eqnarray}
k\,T=\left(\frac{N}{V}\right)^{2/3}\,\frac{h^2}{2\,\pi\,m} \ .
\end{eqnarray}
Considering an ideal classical gas, if we increase the number $N$
of one unity adding, at constant energy and volume, one particle
with kinetic energy zero and neglecting all interactions, the
number of accessible microstates increases and entropy increases
too. Considering the relation
\begin{eqnarray}
\Delta U=T\,\Delta S-P\,\Delta V+\mu\,\Delta N \ ,\label{38}
\end{eqnarray}
we obtain that $\mu$ must be negative. On the contrary, if
repulsive interactions are active, the incoming particle
increases both the energy of the system and the entropy; the
system must spend some energy to return to the initial energy
value. Then $S_{\rm B}$ decreases and sometimes decreases more
than the gained quantity because of one more particle in the
system. In this case $\mu$ is positive.\\ Let us now take for
$\Delta\Omega_0$ the definition of Eq. (\ref{22}). We find that
for an extensive ideal classical gas we have $\mu=0$. This result
can be explained because adding one particle at constant energy
and volume the elementary cells decrease their single volumes and
the total work done by the system to diminish the spatial volume
of $N+1$ cells is $k\,T$. Therefore, from Eq.(\ref{38}) we have
\begin{eqnarray}
\mu={3\over2}\,k\,T-{5\over2}\,k\,T+k\,T=0 \ .
\end{eqnarray}
When the ideal classical gas is nonextensive, using (\ref{311})
the expression of $\mu$ is
\begin{eqnarray}
\mu_\delta=-{45\over4}\,\delta\,k\,T-{25\over2}\,\delta\,N\,k\,T+46\,\delta^2\,k\,T
\ ,\label{40}
\end{eqnarray}
which gives $\mu=0$ when $q\rightarrow1$ ($\delta=0$) and is
slightly positive or negative depending on the sign of $\delta$.
In this case the quantity $\mu$ is composed by intensive and
extensive terms. Its behaviour versus $k\,T$ depends on $\delta$
and on $\delta\,N$ which can be finite or can approach zero or
infinity depending on $\delta$ and $N$, separately. To complete
our study let us discuss the free energy $F$. For an extensive
ideal classical gas with elementary cell $\Delta\Omega_0$ given by
Eq. (\ref{22}) at constant energy and total volume, the function
$F$ and its variation due to the addition of one particle to the
system are given by the following relations:
\begin{eqnarray}
&&F=U-T\,S_{_{\rm B}}=-N\,k\,T \ ,\\ &&\Delta F=-k\,T \ .
\end{eqnarray}
being $\mu=0$  and considering the work spent by the system to
change the spatial volume of elementary cells.\\ The variation of
free energy of a nonextensive ideal classical gas due to the
addition of one particle to the system is, using (\ref{40})
\begin{eqnarray}
\Delta F_q=-P\,\Delta V+\mu_\delta
=-k\,T\,C_\delta+\mu_\delta=-k\,T\,\left(1+{25\over4}\,\delta+{25\over2}\,\delta\,N\right)
\ .\label{43}
\end{eqnarray}
This expression differs from that of the non normalized treatment
of Ref. \cite{23}. The NETS free energy (as also defined in Abe et
al. [18] and with $S_q^{\rm T}$ of  Eq.(\ref{34}) is
\begin{eqnarray}
F_q=U_q-\frac{S_q^{\rm T}}{k_{\rm
T}\,\beta}=-N\,k\,T\,\left(1+{25\over4}\,\delta\,N\right) \
,\label{44}
\end{eqnarray}
and $\Delta\,F_q$ equals the same quantity calculated above [Eq.
(\ref{43})]. A new  definition of free energy introduced by Abe et
al. \cite{18} which is
\begin{eqnarray}
F_q=U_q-T\,{k\over2\,\delta}\,\ln
C_q=U_q-T\,{k\over2\,\delta}\,\ln\left(1+\frac{1-q}{k_{\rm
T}}\,S_q\right) \ ,
\end{eqnarray}
gives, by using the elementary cell we have proposed, the same
result obtained in Eq. (\ref{44}) and then $\Delta F_q$ of Eq.
(\ref{43}).
%%%%%%%%%%%%%%%%%%%%%%%%%%%%%%%%%%%%%%%%%%%%%%%%%%%%%%%%%%%%%%%%%%%%%%

\sect{Interpretation of the parameter $\symbol{113}$}

Let us consider again the square fractional deviation from a MB
distribution
\begin{eqnarray}
\sum_i\frac{(\partial
n_i)^2}{n_{i_0}}=\sum\frac{n_i^2}{n_{i_0}}-N=\left(18\,\delta-\frac{3799}{32}\,\delta^2\right)\,N
\ ,\label{41}
\end{eqnarray}
We have already  discussed that appreciable fractional deviations
are, in general, very small. Nevertheless, few islands of phase
space can give a non negligible contribution to deviation of
quantities calculated within the standard space, like, for
instance, rates of nuclear reactions or atomic processes or rates
of chemical reactions. We define, as usual, the radial correlation
function
\begin{eqnarray}
g(r)=\frac{V}{N}\,\sum_{i\not=j}\delta(\overline{r}+\overline{r}_i-\overline{r}_j)
\ ,
\end{eqnarray}
($g(r)=1$: perfect gas).\\ By using the relation
\begin{eqnarray}
\sum_i\frac{(\partial
n_i)^2}{n_{i_0}}=1+\frac{N}{V}\,\int\left[g(r)-1\right]\,dr \ ,
\end{eqnarray}
from Eq. (\ref{41}) we can write the following relation:
\begin{eqnarray}
18\,\frac{1-q}{2}-\frac{3799}{32}\,\frac{(1-q)^2}{4}={1\over
N}+{1\over V}\,\int\left[g(r)-1\right]\,dr \ .
\end{eqnarray}
Disregarding, for simplicity, the term in $(1-q)^2$ we have
\begin{eqnarray}
q=1-{1\over9}\,\left\{{1\over N}+{1\over
V}\,\int\left[g(r)-1\right]\,dr\right\} \ .
\end{eqnarray}
Therefore, $q\rightarrow1$ if both the conditions $g(r)\rightarrow
1$ and $N\rightarrow\infty$ are fulfilled. Otherwise, if
$N\rightarrow\infty$ but the system is not a perfect gas
$[g(r)\not=1]$:
\begin{eqnarray}
q=1-{1\over9}\,{1\over V}\,\int\left[g(r)-1\right]\,dr \ ,
\end{eqnarray}
and if the system is a perfect gas $(g=1)$  but the number of
particles is small and finite:
\begin{eqnarray}
q=1-{1\over 9\,N} \ .
\end{eqnarray}
Finally, in the limit of a gas/fluid of $N$ particles, we have
found a physical interpretation of the nonextensive parameter $q$
in terms of the particle correlation function among others (see
Wilk et al \cite{6}, Beck \cite{7}, Beck and Cohen \cite{8} and
Tsallis \cite{26}). It can be considered valid at least for values
of $q$  not too different from unity, (small deviations from MB
distribution).

%%%%%%%%%%%%%%%%%%%%%%%%%%%%%%%%%%%%%%%%%%%%%%%%%%%%%%%%%%%%%%%%%%%%%

\sect{Conclusions}

We have calculated the phase space volume $\Omega$ and the
corresponding elementary cell $\Delta\Omega$, occupied by $N$
classical, identical particles composing a weakly nonextensive
system, at equilibrium (or in a steady state, or long-term
stationary state of a nonequilibrium system). Both quantities
must differ from their extensive MB values $\Omega_0$ and
$\Delta\Omega_0$ because we require that the number $W_0$ of
available microstates in the MB extensive phase space $\Omega_0$
equals the number $W$ in the deformed (nonextensive) space:
$W_0=W$. This condition imposes particular analytical expressions
of $\Delta\Omega_0$ and $\Delta\Omega$. The number of elementary
cells results equal to the number of particles $N$.\\ The two
elementary cells result to be macrocells if compared to the
usually taken elementary cell volume, equal to the third power of
the Planck constant , with consequences on the expression of
thermodynamic quantities depending on elementary cell, like, for
instance, the partition function, constant of entropy, chemical
potential and free energy.\\ We have applied results on
elementary cell to nonextensive classical gas, recovering known
results (equation of state) in NETS literature where
$\Delta\Omega$ is usually taken equal to $h^3$ or left
undetermined.\\ We have derived expressions of entropy, chemical
potential and free energy and discussed more deeply the case of
chemical potential. Applications to particular systems and
physical cases like nuclear fusion reactions in plasmas and
chemical reactions will be reported elsewhere. Through
expressions of $\Omega$ and $\Delta\Omega$ we have calculated
fractional deviation from Maxwell-Boltzmann distribution. A
physical meaning of the nonextensive Tsallis parameter $q$ in
terms of the correlation function has been derived.\\

We wish to tank Dr. A.M. Scarfone for critical reading of the
manuscript and comments.
%%%%%%%%%%%%%%%%%%%%%%%%%%%%%%%%%%%%%%%%%%%%%%%%%%%%%%%%%%%%%%%%%%%%%

\vfill\eject
\end{document}